\documentclass{winnower}

\usepackage{graphicx}

\usepackage{amssymb}
\usepackage{natbib}
\usepackage{leftidx}
\usepackage{amsmath}
\usepackage[usenames, dvipsnames]{color}
\usepackage{soul} 
\usepackage{todonotes} 

\newcommand{\x}{\mathbf{x}}
\newcommand{\va}{v^{(1)}}
\newcommand{\vb}{v^{(2)}}

\usepackage[printwatermark]{xwatermark}
\usepackage{xcolor}
\usepackage{graphicx}
\usepackage{lipsum}
\begin{document}

\title{Robust and sparse k-means clustering for high-dimensional data}

\author{Sarka Brodinova}
\affil{Institute of Statistics and Mathematical Methods in Economics, Vienna University of Technology, Austria}
\author{Peter Filzmoser}
\affil{Institute of Statistics and Mathematical Methods in Economics, Vienna University of Technology, Austria}
\author{Thomas Ortner}
\affil{Institute of Statistics and Mathematical Methods in Economics, Vienna University of Technology, Austria}
\author{Christian Breiteneder}
\affil{Institute of Statistics and Mathematical Methods in Economics, Vienna University of Technology, Austria}
\author{Maia Zaharieva}
\affil{Institute of Statistics and Mathematical Methods in Economics, Vienna University of Technology, Austria}

\date{}

\maketitle

\begin{abstract}
\sloppy In real-world application scenarios, the identification of groups poses a significant challenge due to possibly occurring outliers and existing noise variables. Therefore, there is a need for a clustering method which is capable of revealing the group structure in data containing both outliers and noise variables without any pre-knowledge. In this paper, we propose a $k$-means-based algorithm incorporating a weighting function which leads to an automatic weight assignment for each observation. In order to cope with noise variables, a lasso-type penalty is used in an objective function adjusted by observation weights.  We finally introduce a framework for selecting both the number of clusters and variables based on a modified gap statistic. The conducted experiments on simulated and real-world data demonstrate the advantage of the method to identify groups, outliers, and informative variables simultaneously.
\end{abstract}

\section{Introduction}\label{sec:introduction}

\sloppy The identification of groups in real-world high-dimensional datasets reveals challenges due to several aspects: 1) the presence of outliers; 2) the presence of noise variables; 3) the selection of proper parameters for the clustering procedure, e.g. the number of clusters. Whereas we have found a lot of work addressing the three aspects separately, a much smaller number of studies is available in case all three aspects are treated simultaneously. Indeed, in any large and high-dimensional complex dataset, not only outliers but also noise variables are very likely to appear. Hence, a clustering method needs to be designed in such a way that both aspects are taken into account, no matter if outliers are considered as highly interesting observations due to their typically different content or just as noise. The data complexity in terms of the number of groups and the proportion of outliers as well as the number of noise variables very much depends on the dataset itself. Therefore, a clustering procedure should ideally be data-independent. In other words, no information about the data complexity should be assumed. The goal of this paper is to introduce a clustering method designed for such an application scenario.

Considering the task of revealing the group structure in contaminated data, i.e. data with outliers, a natural step is to first apply an outlier detection procedure to exclude deviating observations for the following cluster analysis. However, coping with outliers in such a way might be complicated due to the parameter specification, which is commonly required by most existing clustering (e.g. the number of clusters) as well as outlier detection methods \citep{aggarwal2013}. A better alternative is to use a clustering method which directly incorporates a measure of outlyingness through data clustering in order to reveal clusters and outliers simultaneously as proposed by \citet{campello2015hierarchical}. Another possibility to deal with outliers in the context of clustering is to exclude a certain proportion of deviating observations while applying a clustering method. The idea of excluding observations, which usually do not fit to an assumed model, lead to so-called trimming-based clustering approaches. An overview of such methods can be found in the review by \citet{garcia2010review}. In order to apply a trimming concept, not only the number of clusters but also the trimming level, i.e. the proportion of observations supposed to be discarded, need to be specified in advance. Although \citet{GarcíaEscudero2011}  introduce a diagnostic plot for selecting both parameters using classification trimmed likelihood curves, the procedure depends on the choice of a data-dependent parameter that controls the way how potential outliers should be handled.  Determining such a parameter might however be again difficult for real-world data.

\sloppy The problem of data clustering in the presence of noise variables is usually addressed by sparse- and variable selection-based clustering approaches. The methods generally aim at removing noise variables that can easily mask a group structure \citep{gordon1999classification}. An overview of such methods can be found in the study by \citet{galimberti2017modelling} with a special focus on model-based clustering. Although the number of clusters in model-based clustering is commonly estimated based on the Bayesian information criterion, some methods usually assume that the size of a group is typically larger than the dimensionality of the data space where a group is located.  Therefore, such approaches 
might have troubles to sufficiently discover high-dimensional low sample size groups. A suitable method for such a situation is introduced by \citet{witten2010framework}. The method imposes a lasso-type penalty on incorporated variable weights in the objective function of $k$-means leading to the sparse $k$-means algorithm. In order to apply the sparse $k$-means, the number of clusters needs to be determined in advance, which is hardly possible for most real-world application scenarios. 

The task of identifying groups becomes even more problematic when both outliers and noise variables are present. For this situation, \citet{Kondo2016} introduce the robust and sparse $k$-means (RSKC) that robustifies the sparse k-means by \citet{witten2010framework} by incorporating a trimming concept. However, the approach assumes prior knowledge about the number of clusters, the degree of sparsity, and the trimming level in order to correctly detect clusters. Furthermore, the method has been tested only in terms of clustering and no evaluation has been performed regarding the detection of outliers. Such observations  may additionally provide useful information about the analyzed datasets since they usually differ from the main group structure. 

In contrast to RSKC, we introduce a robust and sparse $k$-means-based procedure that is capable of finding the true underlying structure in very complex data, i.e. data containing clusters, outliers, and noise variables simultaneously.  The presented $k$-means-based algorithm incorporates a weighting function employing a measure of outlyingness in order to automatically assign a weight to each observation. While a high weight indicates that  an observation is part of a cluster,  a low weight  refers to a potential outlier. The advantage of using a weighting function  is that we do not have to pre-specify any trimming level as for trimming-based approaches. To exclude noise variables, we use a lasso-type penalty imposed on the variable weights in an objective function adjusted by observation weights. In order to correctly detect groups, we eventually propose a framework aiming at the determination of the optimal parameters, such as the degree of sparsity and the number of clusters.

The rest of this paper is organized as follows. Section~\ref{sec:algorithm} briefly reviews $k$-means-based clustering approaches and motivates the proposed clustering procedure which is described in detail in Sect.~\ref{sec:method}. The parameter selection is presented in Sect.~\ref{sec:parameters} and 
thoroughly tested on simulated data sets in Sect.~\ref{sec:simulation}.  We compare the proposed method with other $k$-means-based clustering methods on a real-world dataset in Sect.~\ref{sec:application}.  Section~\ref{sec:discussion} concludes the paper.

\section{$k$-means-based algorithms}\label{sec:algorithm}

Despite the large number of developed clustering procedures, $k$-means remains one of the most popular and simplest partition algorithms \citep{jain2010data}. Given a data matrix $\mathbf{X}=\{ x_{ij} \}, i=1,\dots,n, \, j=1,\dots,p,$  with $n$ observations described by $p$ variables, the task of finding $k$ clusters based on $k$-means  was originally established using the within-cluster sum of squares $W^k$ for the given number of clusters $k$ as 
\begin{equation}\label{eq:kmeans1}
\begin{split}
W^k_j&= \sum_{r=1}^k \sum_{i \in K_r} (x_{ij} -  m_{jr})^2,\\
 W^k &=\sum_{j=1}^p W^k_j
\rightarrow  \min_{K_1, \dots, K_k},
\end{split}
\end{equation}
where $W^k_j$ corresponds to the within-cluster sum of squares in the $j^{th}$ variable and the set $K_r$~contains the indices of the observations assigned to the $r^{th}$ cluster, for $r=1,\ldots ,k$. Note that such an optimization problem can also be reformulated with respect to the between-cluster sum of squares  $B^k$ \citep{witten2010framework} as
\begin{equation}\label{eq:kmeans}
\begin{split}
B^k_j&=\sum_{i=1}^n  (x_{ij} - m_j)^2 - \sum_{r=1}^k \sum_{i\in K_r} (x_{ij} -  m_{jr})^2,  \\
B^k &=\sum_{j=1}^p B^k_j \rightarrow  \max_{K_1, \dots, K_k},
\end{split}
\end{equation}
where $B^k_j$ denotes $B^k$ in the $j^{th}$ variable,  $m_j$ is the $j^{th}$ coordinate of the overall data center, and $m_{jr}$ denotes the center of the $r^{th}$ cluster in the $j^{th}$ variable. 

Although $k$-means is very popular, it  has several disadvantages that need to be taken into account when developing a clustering procedure. The first drawback of $k$-means is  the random initialization of cluster centers, which may lead to non-optimal solutions. This can be overcome by using an appropriate initialization method; an overview of such approaches can be found in a study by \citet{celebi2013comparative}. For our method, we incorporate the ROBIN (ROBust INitialization) approach by \citet{Hasan2009}. The method is able to find optimal centers in a small number of runs unlike the original $k$-means. ROBIN seeks for initial centers that are located in the most dense region and are simultaneously far away from each other in order to avoid the selection of outliers as initial centers. In order to identify the observations in highly dense regions, ROBIN uses LOF (Local Outlier Factor) proposed by \citet{Breunig2000}. LOF was primarily introduced to measure a degree of outlyingness of observations in complex data where observations tend to form groups. The method compares local densities of observations with the local densities of their $q$ nearest neighbors using various ratios of the Euclidean distances. The resulting outlyingness, $lof_q(\mathbf{x}_i)$, of an observation $\mathbf{x}_i$ close to 1 indicates that $\mathbf{x}_i$ is potentially part of a cluster and, therefore, a candidate for an initial cluster center, as proposed by ROBIN. In contrast, $lof_q(\mathbf{x}_i)\gg 1$ suggests that $\x_i$ is a possible outlier and thus $\x_i$ should not be considered as an initial center.

The second limitation of $k$-means is the employed sample mean that suffers from a lack of robustness. As a result, $k$-means is also not resistant against outliers and even a single deviating observation can affect the final clustering solution \citep{garcia1999}. In order to robustify $k$-means, \cite{cuesta1997trimmed} proposed a trimmed version defined as
\begin{equation}\label{eq:kmeafns}
\begin{split}
{^t}B^k_j& = \sum_{i\in L}  (x_{ij} - m_j)^2 - \sum_{r=1}^k \sum_{i \in K_r \cap L} (x_{ij} -  m_{jr})^2,  \\
 {^t} B^k & =\sum_{j=1}^p { ^t} B^k_j \rightarrow  \max_{K_1, \dots, K_k, L},
\end{split}
\end{equation}
where ${^t} B^k  =\sum_{j=1}^p { ^t} B^k_j$ represents the between-cluster sum of squares calculated on the untrimmed observations, $L$ denotes the set containing indices of $[n(1-\alpha)]$ (untrimmed) observations that have the smallest distance to their closest cluster center, and $\alpha$ is the trimming level. Such a robustification excludes the $\alpha$ fraction of  observations, i.e. potential outliers,  for calculating the cluster centers in order to achieve an accurate clustering solution if $\alpha$ is chosen correctly according to the true outlier proportion. Determining $\alpha$ may however be problematic for real-world data.  In order to avoid the parameter-dependent robust $k$-means, we propose to incorporate a measurement of outlyingness which leads to a clear decision on determining outliers. Such a concept was introduced by \citet{Filzmoser2008} in case of a one group data structure resulting in observation weights. The weights reflect how much an observation is outlying on the $[0,1]$-scale with a low weight indicating a potential outlier. We incorporate the concept of observation weights in $k$-means in order to robustify the method in such a way that no parameter pre-specification is required.

The last disadvantage of $k$-means occurs when a group structure is detectable only in a small subset of variables. In order to find such variables, \citet{witten2010framework} introduced a framework for sparse $k$-means based on a lasso-type penalty leading to the problem of maximizing the weighted $B^k$ for a given $k$ and a sparsity parameter $s$ as
\begin{equation}\label{wbcss}
B^{sk}=\sum_{j=1}^p w_j B^k_j
\rightarrow \max_{K_1, \dots, K_k, \mathbf{w}}, 
\end{equation}
subject to $||\mathbf{w}||^2 \leq 1, ||\mathbf{w}||_1 \leq s$ for  $\mathbf{w}=\{w_j \ge 0 \} \,\forall j$ and $s \in(1, \sqrt{p}\,]$, which can be solved in an iterative way  as proposed by \citet{witten2010framework}. The parameter $s$ controls the degree of sparsity in the variable weight vector, i.e. the values of $w_j$. The more important (informative) the $j^{th}$ variable, the higher the value of $w_j$. Our method also uses a lasso-type penalty in the objective function, but the value of $B^{sk}$ is additionally adjusted by observation weights in order to achieve robustness. Although, the proposed method is similar to RSKC by \citet{Kondo2016}, our procedure can be seen as a better alternative since no trimming level is required.  In addition, we aim at analyzing the data structure more thoroughly, i.e. discovering clusters, outliers, and informative variables simultaneously.

\section{The proposed algorithm}\label{sec:method}

The introduced method is an iterative three-step approach. In the first step, $k$-means employing a weighting function is applied on the data space spanned by the variables with some contribution to a cluster separation (i.e. with the variables having $w_j>0$, see~Eq.~(\ref{wbcss})). The incorporated weighting function robustifies $k$-means and results in observation weights reflecting the outlyingness. The second step aims at updating the variable weights with respect to both clusters and observation weights from the first step. The two steps are iteratively repeated until the variable weights stabilize. In the third step, the observations are clustered with respect to the identified informative variables and the observations with small weights are classified as outliers. The detailed description of the algorithm is given in the following subsections.

\subsection{Step 1: Downweighting outlying observations}\label{sec:out}

The aim of the first step is to robustify $k$-means by  incorporating a weighing function in order to downweight the influence of potential outliers. Assuming that the number of clusters $k$ is known, we apply ROBIN   with $q=10$ \citep{Hasan2009} on weighted data, $_w\mathbf{X}= \{ _w \mathbf{x}_i\} =\{w_j \, x_{ij} \},  \forall i,  \forall j,$ where $\mathbf{w}=\{ w_j = 1/\sqrt{p}\}, \forall j$, in order to find the first $k$ cluster centers. Note that these initial values for $w_j$ are considered only in the first iteration as recommended by \citet{witten2010framework}, but in the next iteration $w_j$ will be already different and will better reflect the contribution to a cluster separation.  

After applying ROBIN, each observation is assigned to its closest cluster center leading to the corresponding cluster membership $K_1, \dots, K_k$. We then propose to apply a weighting function on the detected clusters to reveal outliers. The weighting function should be a monotonic decreasing function using an outlyingness measure as an argument in order to obtain observation weights that range between 0 and 1, with a low weight indicating a potential outlier.  Hence, it is essential to choose both a suitable outlyingness measure and an appropriate weighting function.

A naive approach is to calculate the Euclidean distance of an observation to its closest cluster center. However, using the Euclidean distance provides the information about how far an observation is from its closest center rather than how much an observation deviates or to what degree it is an outlier. In fact, such information can be easily obtained by applying LOF on each detected cluster as $lof(_w\x_i):= lof_q(_w \x_i), i\in K_r, \, \forall r$, with $q=10$ as recommended by \citet{Breunig2000, Hasan2009}.
The LOF scores are then standardized as
\begin{equation}\label{eq:lof}
 lof^*_i = \frac{lof(_w\x_i) - \text{mean}(lof(_w\x_i),i\in K_r)}{\text{sd}(lof(_w\x_i),i \in K_r)} 
\end{equation}
to be suitable for the weighting function with the mentioned properties. Preliminary studies indicated good empirical results when the observation weights, denoted as $\va_i$, were obtained using the translated bi-weight function \citep{Rocke1996} as follows
\begin{equation}\label{eq:v}
    \va_i=
    \begin{cases}
      0, &  lof^*_i \ge c \\
     \Bigg( 1- \Big ( \frac{lof^*_i - M}{c-M} \Big)^2 \Bigg )^2, & M <  lof^*_i <c, \\
    1, &  lof^*_i \leq M
    \end{cases}
 \end{equation}
\sloppy where $i\in K_r, \forall r$, $M=\text{med}(lof^*_i,i\in K_r) + \text{MAD}( lof^*_i, i\in K_r)$, and $c=2$. The obtained weights correspond to the measure of outlyingness values in $[0,1]$. While a value close to 1 indicates that an observation is part of a cluster, $\va_i \approx 0$ suggests that $\x_i$ is an outlier with respect to the detected cluster. The weights based on LOF better express the degree of deviation than e.g. using a simple Euclidean distance of an observation to the closest cluster as in RSKC. In addition, the weights should be more robust against elliptically-shaped clusters due to the properties of LOF; see \citet{Breunig2000}. If the shape of a cluster is slightly elliptical, RSKC might exclude observations which are further away from the cluster center but still part of a cluster.

After assigning weights to observations from each detected cluster according to (\ref{eq:lof}) and (\ref{eq:v}), we plug the weights $\va_i$ into the weighted between-cluster sum of squares for a given $\mathbf{w}$, and optimize the cluster assignment as 
\begin{equation}\label{eq:aj_our}
\begin{split}
^{\va} B^k_j & = \sum_{i=1}^n   \va_i \Big( x_{ij} - \frac{1}{\sum_{i=1}^n \va_i} \sum_{i=1}^n \va_i x_{ij} \Big )^2 \\
& - \sum_{r=1}^k \sum_{i\in K_r} \va_i  \Big( x_{ij} -  \frac{1}{\sum_{i \in K_r} \va_i} \sum_{i \in K_r} \va_i  x_{ij} \Big)^2
\end{split}
\end{equation}
\begin{equation}\label{eq:aj_max}
 \sum_{j=1}^p w_j \,^{\va} B^k_j
\rightarrow  \max_{K_1, \dots, K_k} 
\end{equation}
in order to robustify $k$-means. We can clearly see from (\ref{eq:aj_our}) that if an observation is  a potential outlier, i.e $ \va_i \approx 0$, the distance of such an observation to its closest cluster center is downweighted by the corresponding value of $\va_i$. In contrast, an observation with $\va_i \approx 1$ highly contributes to the maximization. The observation weights are also used to determine the next cluster centers, i.e. $\frac{1}{\sum_{i \in K_r} \va_i} \sum_{i \in K_r} \va_i  x_{ij}, \forall r$, in a robust way by using the weighted mean of observations in each coordinate. 
The cluster centers with the corresponding cluster assignment are iteratively updated until a local optimum is reached during a certain number of iterations in the sense of maximization of~(\ref{eq:aj_max}).  In our experiments the method is allowed to search for the local optimum during 15 iterations, but also a higher number can be considered. Note that the local optimum is achieved  on the weighted data, i.e. in a data space spanned by the variable vector with $w_j>0$ adjusted by the values of $w_j$. 

\begin{figure}[h]
\begin{center}
    \includegraphics[width=0.5\textwidth]{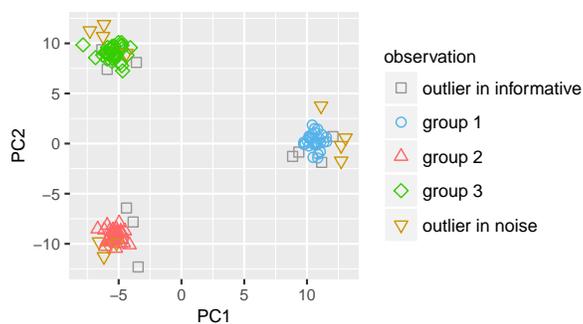}
\caption{The generated dataset shown in the principal component space. The observations from 3 groups, outliers placed in informative and noise variables are displayed in different colors and symbols}\label{Fig1}
\end{center}
\end{figure}

\begin{figure}[h]
\begin{center}
    \includegraphics[width=1\textwidth]{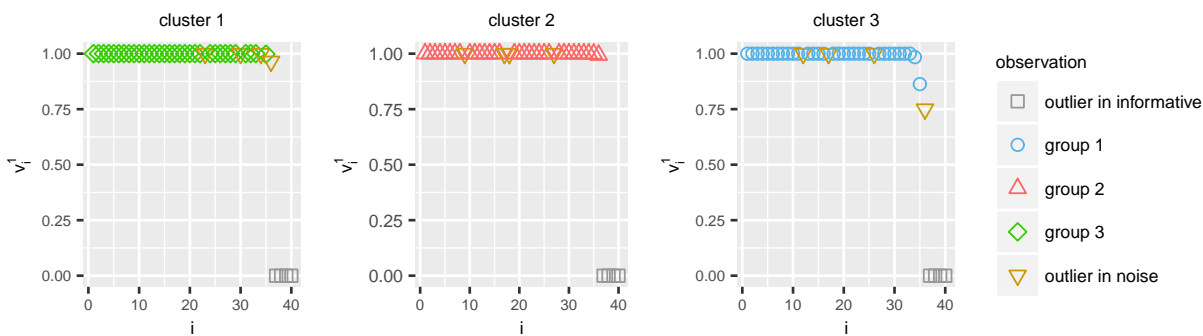}
\caption{Illustration of incorporating the weighting function in $k$-means in order to reveal outliers as observations with low $\va_i$ and to detect 3 clusters on the weighted data}\label{Fig2}
\end{center}
\end{figure}

We illustrate the efficiency of the weighting function on an example dataset that consists of three groups with the same sizes of 40 observations. The group structure is described by 50 variables leading to  high-dimensional low sample size groups. We add 750 noise variables and contaminate $10\%$ of the observations from each group in the informative variables and in 75 noise variables; a detailed description of the data setup is provided in Sect. \ref{sec:simulation} and corresponds to the first simulation study. Figure~\ref{Fig1} visualizes the generated dataset in the space spanned by the first two principal components; the group membership and outliers are differentiated by colors and symbols.
The final weights, obtained during two iterations given the initial cluster centers by ROBIN, are shown in Fig.~\ref{Fig2} in decreasing order to visualize the shape of the weighting function.  Importantly, the observation weights are calculated in the data space defined by 50 informative variables. In other words, we now assume that $\mathbf{w}$ is known beforehand in order to demonstrate the concept of the weighting function. We can see in Fig.~\ref{Fig2} that all observations from group~3 are correctly assigned to cluster~1 because no observations from group 3 are visible in the following plots. The plot particularly indicates that the weighting function works properly since all non-outliers obtain a weight around 1. In contrast, outliers placed in informative variables receive a weight around 0 and can thus be easily identified. A similar conclusion can be made in case of the other two clusters. The plots may suggest that the non-outliers with a weight smaller than 1 could be located on the edge of a cluster or slightly further from the other clustered observations. However, we cannot reveal outliers placed in noise variables as indicated by their weights equal to 1, since the noise variables are not involved in the clustering due to their zero weights.

\begin{figure}[h]
\begin{center}
    \includegraphics[width=1\textwidth]{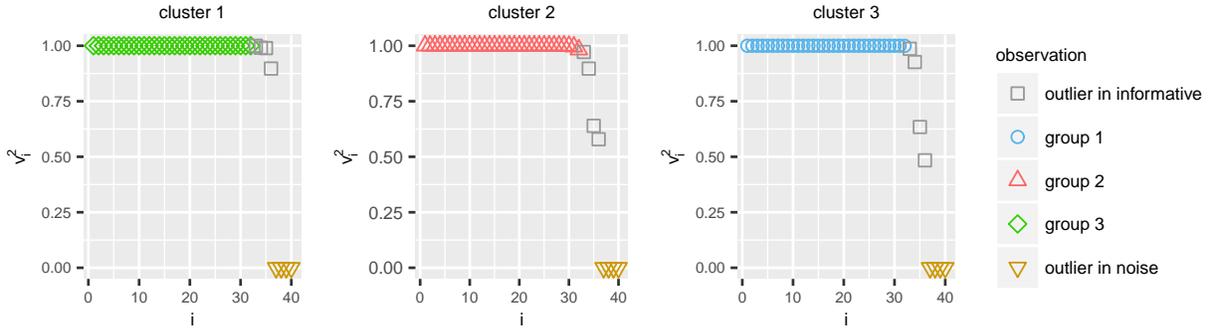}
\caption{Illustration of applying the weighting function on the 3 unweighted data clusters in order to reveal outliers in noise variables as observations with low $\vb_i$}\label{Fig3}
\end{center}
\end{figure}

In order to identify outliers in noise variables as well, we additionally apply the proposed weighting function on unweighted data clusters, consisting of the data matrices $\mathbf{X}_r=\{ x_{ij}\}, i\in K_r, \, \forall j=1, r=1,\dots, k$, leading to the second observation weights $\vb_i$. Figure \ref{Fig3} shows the second resulting observation weights obtained on the data example shown in Fig. \ref{Fig1}. The three plots clearly indicate that all outliers placed in noise variables receive considerably lower weights  in contrast to both non-outliers and outliers present in informative variables. 

\begin{figure}[h]
\begin{center}
    \includegraphics[width=1\textwidth]{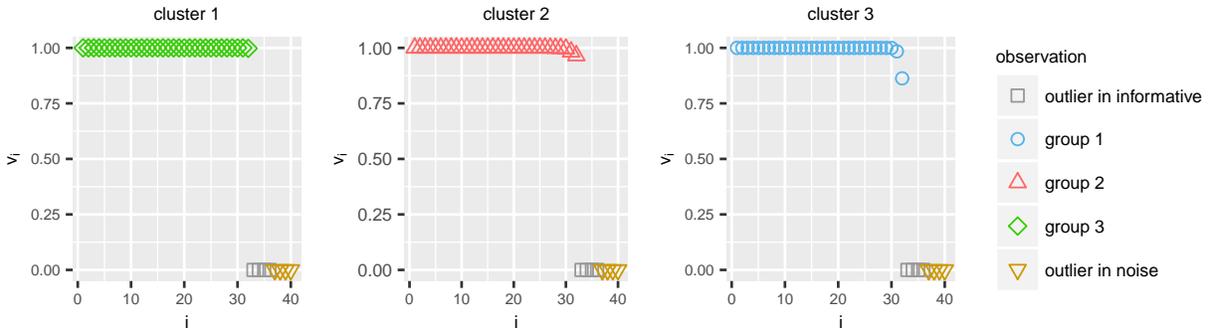}
\caption{
Illustration of combining the two observations weights leading to the final observation weights $v_i$ calculated on 3 clusters}\label{Fig4}
\end{center}
\end{figure}

As a consequence of applying the weighting function for the second time, each observation has two weights, $\va_i$ and $\vb_i$, which are finally combined in a single weight 
\begin{equation}
v_i =\min \{ \va_i, \vb_i \}.
\end{equation}
Determining $v_i$ in this way ensures that all outliers receive low weights and that we can easily identify whether or not an observation is an outlier as indicated by zero weights for all outliers in Fig. \ref{Fig4}.

\subsection{Step 2: Variable selection}

The purpose of the second step is to update $w_j$ according to the maximization of (\ref{wbcss}) for a given sparsity parameter $s$ and the observation weights $v_i$. Incorporating $v_i$ assures that the variable selection is not affected by outliers. Indeed, the presence of an outlier apparent even in one variable can considerably increase the between-cluster sum of squares. As a result, such a variable receives a high weight although the variable does not contribute to the cluster separation but rather to the separation between an outlier and a cluster \citep{Kondo2016}.

Therefore, for the obtained cluster assignment $K_1$,\dots, $K_k$, the observation weights $v_i$ from the first step, and for a given $s$, we update the weights $w_j$  according to
\begin{equation}\label{eq:aj}
^{v} B^{ks} = \sum_{j=1}^p w_j \, ^{v}B^k_j
\rightarrow \max_{ ||\mathbf{w}||^2 \leq 1, ||\mathbf{w}||_1 \leq s}, 
\end{equation}
where $^{v} B^k_j$ corresponds to Eq. (\ref{eq:aj_our}) with $v_i$ instead. In order to optimize (\ref{eq:aj}) with respect to $\mathbf{w}$ for a given tuning parameter $s$, we follow the procedure suggested by \citet{witten2010framework}. Whereas small $s$ leads to high sparsity, i.e. $w_j=0$ for most variables, a high value of $s$ results in almost no sparsity corresponding to $w_j>0$ for most variables. High $w_j$ suggests that the $j^{th}$ variable is informative and, thus, it contributes to the maximization of (\ref{eq:aj}). In contrast, $w_j=0$ indicates that the $j^{th}$ variable is not informative for the cluster separation and it is thus excluded in (\ref{eq:aj}).

Once the variable weights $\mathbf{w}$ are updated, the first iteration is completed and the algorithm continues with the first step with respect to updated weights $w_j$. This means that the ROBIN approach is again applied on $_w\mathbf{X}= \{ _w \mathbf{x}_i\}$ with updated $\mathbf{w}$ in order to find the next cluster centers. The reason for the re-initialization is that ROBIN is not primarily designed to deal with a large number of noise variables. Therefore, the selection of the first cluster centers is very likely to be affected by noise variables due to $\mathbf{w} =\{ w_j=1/\sqrt{p}\}, \forall j$ in the first step. After obtaining the next centers, the method continues as described. The two steps of the proposed approach are iteratively repeated until convergence for $w_j$ is reached according to the stopping criterion \citep{witten2010framework}. 

\subsection{Step 3: Detection of groups and outliers}\label{seq:outgroup}

The last step aims at determining the cluster membership $K_1, \dots, K_k$ by assigning observations to their closest cluster center in the data space spanned by variables with $w_j>0$, adjusted by their corresponding  final weights. We estimate the final observation weights $v_i$ as described in Sect. \ref{sec:out} in order to classify observations with low weights as outliers. This classification can be made based on visualization of the resulting observation weights against the corresponding observation index, as shown in Fig. \ref{Fig4}, and the following search for a cut-off value which clearly separates low weights from high weights. Nevertheless, we recommend to use $v_i<0.5$ for the identification of outliers as we observed good empirical results for such a choice.

\section{Selection of parameters}\label{sec:parameters}

We have so far assumed pre-knowledge about the number of clusters $k$ as well as the tuning parameter $s$ determining the variable weights $w_j$. Such information is usually not available beforehand for most real-world data and, therefore, there is a need for a systematic way of estimating both parameters. The problem of selecting the optimal $k$ has been widely studied for data where the assumption is that all variables are involved in data clustering; an overview of such procedures can be found in the studies by \citet{sugar2003finding} and \citet{xu2005survey}. However, we have not found much work dedicated to the optimization of $k$ in case that the sizes of groups are much lower than the dimensionality of 
the data space describing the group structure and at the same time the group structure is hidden in a large number of noise variables. 

We discuss the effect when $k$ is optimized with and without taking the contribution of variables into account using the gap statistic \citep{tibshirani2001estimating}. The gap statistic, $Gap_k$, is calculated for a clustering solution obtained by a clustering algorithm, e.g. $k$-means, for a given $k$ and can be formulated as
\begin{equation}\label{eq:gapk}
Gap_k= \sum_{j=1}^p w_j \Big( \frac{1}{A} \sum_{a=1}^A \log({_a}W_{j}^{k}) - \log (W_{j}^k) \Big),
\end{equation}
where $w_j=1, \forall j$ since all variables are assumed to contribute equally, ${_a}W^k = \sum_j {_a}W_{j}^k$ corresponds to $W^k= \sum_j W_{j}^k$ calculated on the clustering solution obtained on the dataset with independently permuted observations in each variable \citep{witten2010framework}, and $A$ represents the number of permuted datasets. In our experiments we consider $A=10$. $Gap_k$ is generally calculated for a clustering solution with varying $k$ and the optimal number of clusters is chosen as the smallest $k$ for which $Gap_k \ge Gap_{k+1} - se_{k+1} $ is fulfilled \citep{tibshirani2001estimating}, where $se_k$ denotes the standard error of $\log({_a}W^k)$. From (\ref{eq:gapk}) it is obvious that $Gap_k$ does not only depend on $k$ but also on $\mathbf{w}$ representing the contribution of each variable. Since all variables are assumed to be informative, $Gap_k$ might be considerably affected if a dataset contains a large number of noise variables.  Moreover, the presence of deviating observations can lead to an unreliable decision on $k$ as well.  

\begin{figure}[h]
\begin{center}
    \includegraphics[width=0.95\textwidth]{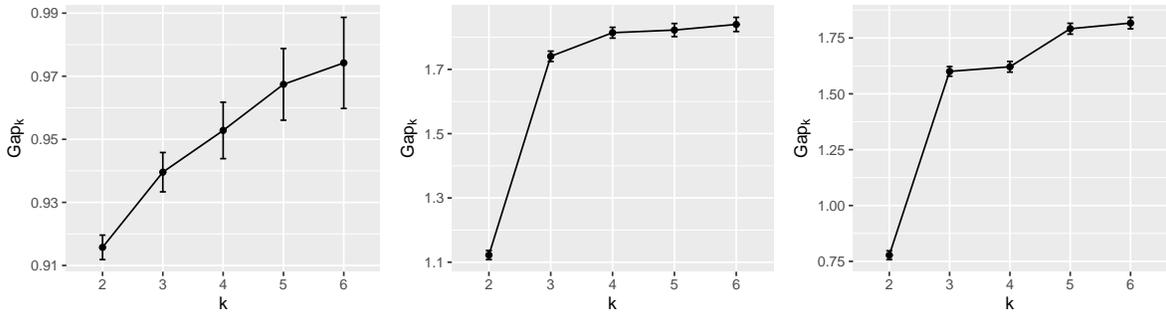}
\caption{The effect of noise variables and outliers when estimating the optimal number of clusters. The values of $Gap_k$ applied on a data set with both noise variables and outliers (left); the resulting $Gap_k$ from a dataset where the effect of noise variables is eliminated (middle); the obtained $Gap_k$ when both noise variables and outliers are neglected (right)}
\label{Fig5}
\end{center}
\end{figure}

Figure \ref{Fig5} demonstrates the effect of noise variables and outliers on the choice of $k$ based on the gap statistic. We consider the same data example as in Sect. \ref{seq:outgroup} and apply $k$-means with ROBIN initialization for  the numbers of clusters $k=2,\dots,6$. The gap statistic is calculated for each clustering solution in order to select the optimal $k$ as described above. Figure \ref{Fig5} (left) shows the values of $Gap_k$ with the corresponding standard errors calculated on the data example with both outliers and noise variables. As expected, the presence of both disturbing factors leads to a wrong choice of the optimal $k$ corresponding to 5 clusters.  Moreover, even if only the 50 informative variables are taken into account, the choice of $k$ is also influenced by outliers as illustrated in Fig. \ref{Fig5} (middle) resulting in $k=4$. In contrast, Figure \ref{Fig5} (right) shows $Gap_k$ when downweighting outlying observations and noise variables leading to a correct decision, i.e $k=3$.

The example indicates that both disturbing factors have to be considered when selecting an optimal $k$ for $k$-means. In the proposed $k$-means-based clustering approach, we directly downweight the effect of outliers by observation weights $v_i$.  However, the impact of noise variables, which is reflected by their corresponding variable weights $w_j$, can be neglected only if the sparsity parameter $s$, see Eq. (\ref{eq:aj}), is correctly selected. In order to optimize $s$, \citet{witten2010framework}  introduced the gap statistic $Gap_s$, which is defined for given $k$ as 
\begin{equation}
Gap_{s} = \log (B^{sk}) - \frac{1}{A} \sum_{a=1}^A \log({_a}B^{sk}),
\end{equation}
where ${_a}B^{sk}$ denotes the weighted between-cluster sum of squares calculated, compare (\ref{wbcss}), with respect to a clustering solution obtained on a permuted dataset. Obviously, the calculation of $Gap_{s}$  is impossible if the number of clusters $k$ is unknown which is often the case for real data. Moreover, the presence of outliers might also influence the correct estimation of $s$. Therefore, we propose to adjust $Gap_{s}$ by observation weights $v_i$ in order to downweight the influence of outliers leading to the modified gap statistic ${^v}Gap_{sk}$ calculated as
\begin{equation}
{^v}Gap_{sk}= \log ({^v}B^{sk}) - \frac{1}{A} \sum_{a=1}^A \log(\leftidx{_a^v}B^{sk}),
\end{equation}
where $\leftidx{_a^v}B^{sk}$ represents ${^v}B^{sk}$ obtained on a permuted dataset. We calculate ${^v}Gap_{sk}$ for a clustering solution not only with various $s$ but also various $k$ in order to first optimize the degree of sparsity $s$ for each $k$. The value of ${^v}Gap_{s^*k}$ for the optimal parameter $s^*$ is compared with the largest ${^v}Gap_{sk}$ such that ${^v}Gap_{s^*k} \ge  {^v}Gap_{sk} - se_{sk}$,  where $se_k$ refers to the standard error of $\log(\leftidx{_a^v}B^{sk})$. The optimization of $s$  leads to $k$ values of ${^v}Gap_{s^*k}$ for which the largest value corresponds to an optimal $k$.

\begin{figure}[h]
\begin{center}
    \includegraphics[width=1\textwidth]{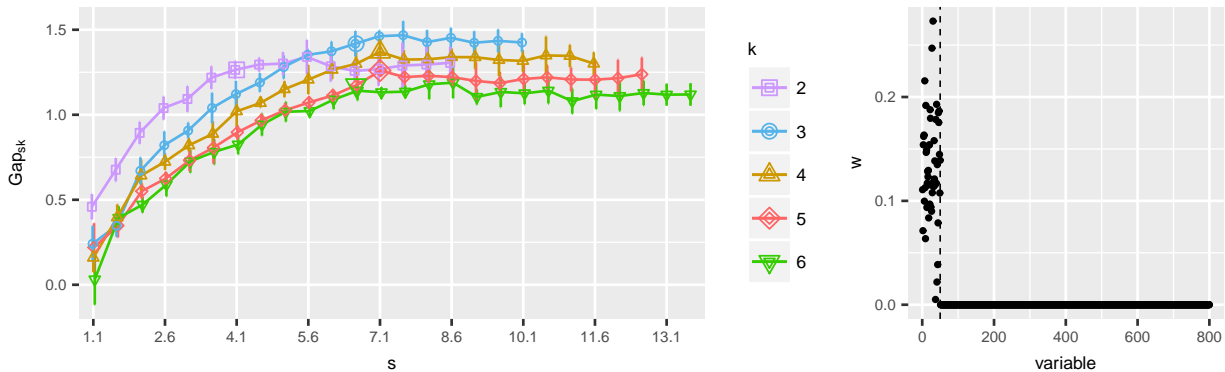}
\caption{ Selection of the tuning parameter $s$ and the number of clusters $k$ based on ${^v}Gap_{sk}$ (left), and the variable weights corresponding to the optimal $s=6.6$ and $k=3$ (right)}\label{Fig6}
\end{center}
\end{figure}

Figure \ref{Fig6} (left) depicts the gap statistic for both tuning parameters when applying the proposed method on the data example in Sect.~\ref{seq:outgroup} with $k=2,\dots,6$. The value of $s$ starts at 1.1 and increases in steps of 0.5 to such a value that leads to no sparsity in the variable weights, i.e $w_j \neq 0, \forall j$. We show the optimal $s$ for each $k$ by larger symbols in Fig.~\ref{Fig6} (left). As expected, the optimal degree of sparsity $s$ differs almost for all $k$. We select the optimal parameter setting which 
leads to the largest ${^v}Gap_{s^*k}$ resulting in $k=3$ and $s=6.6$. Indeed, such choices correspond to the correct number of clusters as well as appropriate values of $w_j$ leading to  non-zero weights for all 50 informative variables as shown in Fig. \ref{Fig6} (right). Considering higher values of $s$, more and more noise variables obtain non-zero weights. The plot additionally illustrates that a smaller choice of $s$, e.g. $s=4.1$, results in an incorrect number of clusters when following the rule for optimizing $k$ based on $Gap_k$ according to \citet{tibshirani2001estimating}. This supports the fact that both parameters need to be optimized at the same time in order to correctly identify groups.

\section{Evaluation setup}\label{sec:evalsetup}

We evaluate the performance of the proposed method in terms of the clustering solution, outlier detection, and the identification of informative variables. The clustering solution is evaluated based on the Classification Error Rate (CER), also used by \citet{witten2010framework}. CER compares the true group membership with the resulting cluster membership.  While CER=$0$ refers to the best cluster solution, CER=$1$ corresponds to the poorest performance.  
In order to evaluate the ability of our method to detect outliers, we report the mean value of observation weights $v_i$ separately for the true non-outliers, i.e $\bar{v}^{nonout}$, and outliers, denoted as $\bar{v}^{out}$. The weights for outliers are supposed to be considerably lower than the weights for non-outliers. Since we recommend to use the final weights $v_i$  for classifying outliers as the observations with $v_i <0.5$, we calculate True Positive and False Positive Rates (TPR and FPR) ranging between 0 and 1. TPR is defined as the proportion of the number of correctly identified outliers and the actual number of outliers present in a given dataset. High TPR indicates a good ability to identify outliers while low TPR demonstrates poor performance. FPR is calculated as the ratio between the number of non-outliers wrongly declared as outliers  and the number of the actual non-outliers in an analyzed dataset. Hence, low values of FPR are preferable over high values.  
The performance regarding the variable selection is evaluated by comparing the mean value of $w_j$ for informative variables, $\bar{w}^{\,inf}$, with the mean value of $w_j$ that are different from zero, denoted as $\bar{w}^{\,non0}$. The higher and more similar the values, the better the ability to correctly select informative variables. We provide a similar evaluation for noise variables and calculate the mean of their weights, $\bar{w}^{\,noise}$, which is supposed to be close to zero.

Since the clustering procedure employs $k$-means, we compare the method with several existing $k$-means-based clustering algorithms, such as $k$-means~(K)\footnote{We employed the code implemented in the R package \texttt{RSKC} \citep{KondoR}.}, 
trimmed $k$-means~(TKC)\footnotemark[\value{footnote}] by \citet{cuesta1997trimmed}, 
and sparse $k$-means~(SKC)\footnote{The used code for sparse $k$-means as available in the R package \texttt{sparcl}  \citep{sparcl}} by \citet{witten2010framework}. 
The proposed weighted robust and sparse $k$-means~(WRSK) is also compared with trimmed and sparse $k$-means (RSKC)\footnotemark[\value{footnote}]  by \citet{Kondo2016}. 
Although our algorithm is designed in a similar way as RSKC, we avoid to  pre-specify the trimming level by incorporating the proposed weighted function. Since no procedure for selecting the optimal $k$ and $s$ has been presented by \citet{Kondo2016} for RSKC in case that no information about data is available, we employ the modified gap statistic considering zero weights for trimmed observations and weights  equal to one for untrimmed observations. Note that all trimming-based algorithms require for the pre-specification of a trimming level $\alpha$, therefore, when applying these methods we consider $\alpha$ as the true percentage of outliers present in a simulated dataset and $\alpha=0.10$ for real-world data as recommended by \citet{Kondo2016} being a suitable choice for most cases.

\nocite{stats}

\section{Simulation study} \label{sec:simulation}

In this section, we explore the ability of the proposed clustering method to correctly  reveal the complex data structure in three simulation studies. We first show the efficiency of the gap statistic to properly select $s$ and $k$. Then, we test the method on the datasets containing various percentages of outliers.  Finally, the proposed method is compared with several existing $k$-means-based approaches.

We now describe the general setting of the simulated datasets considered in the three studies. Each dataset consists of $n$ observations described by the informative as well as uninformative part in terms of the group separation. The observations in the informative part form $g$ groups of sizes $n_t, t= 1, \dots, g$.  The groups are described by $p_{inf}$ variables and are generated following a Gaussian
model with parameters  $\boldsymbol{\mu}_t \in \mathbb{R}^{ p_{inf}}$  and $\boldsymbol{\Sigma}_t \in \mathbb{R}^{p_{inf} \times p_{inf}}$. The elements of the mean vector $\boldsymbol{\mu}_t=(\mu_{t1},\dots, \mu_{tp_{inf}})$ are constructed as
\begin{equation}
    \mu_{tj}=
    \begin{cases}
      \mu, &   j=a_z,  \\
      0, & \text{else}
    \end{cases}
 \end{equation}
where $\mu$ is randomly chosen from the uniform distribution in $[-6,-3] \cup [3,6]$, i.e $ U[-6,-3] \cup \, U[3,6]$. $a_z$ represents the arithmetic sequence defined as $a_{z+1}=a_z + g, a_1=t$ meaning that the first nonzero element of $\boldsymbol{\mu}_t$ is placed on the $t^{th}$ position and the following nonzero elements, i.e. $\mu$,  are always on the position increased by $g$ with respect to the previous index of the nonzero element.  Considering, for example, 4 groups of 10 dimensions, the  mean vectors of the first two clusters are constructed as $\boldsymbol{\mu}_1=(\mu,0,0,0,\mu,0,0,0,\mu,0,0)$ and $\boldsymbol{\mu}_2=(0,\mu,0,0,0,\mu,0,0,0,\mu,0)$. The covariance matrix $\boldsymbol{\Sigma}_t$ is generated according to \citet{campello2015hierarchical} as 
\begin{flalign}
\boldsymbol{\Sigma}_t= \mathbf{Q} \left( \begin{array}{ccccc}
1 & \rho^t & \dots & \rho^t\\
\rho^t & \ddots & \ddots  & \vdots  \\
\vdots & \ddots  &   \ddots & \rho^t\\
\rho^t &  \dots& \rho^t & 1
\end{array} \right) \mathbf{Q}^\top,
\end{flalign}
where $\mathbf{Q}$ denotes a random rotation matrix satisfying $\mathbf{Q}^\top=\mathbf{Q}^{-1}$ and  the off-diagonal elements $\rho$ are random numbers from $U[0.1,0.9]$.   
To the informative part, we also add $p_{noise}$ noise variables that follow univariate standard normal distributions leading to a total dimensionality of $p=p_{inf}+p_{noise}$.  

Such an obtained dataset is finally contaminated by replacing a certain percentage of observations in each group by outliers. We create two types of outliers in the informative variables. While uniformly distributed outliers are generated as random values from $U[-12,6] \cup U[6,12]$, the scattered outliers follow a Gaussian model with the same location as a group, i.e. $\boldsymbol{\mu}_t $,  but a different covariance structure $\sigma \mathbf{I} \in \mathbb{R}^{p_{inf} \times p_{inf}}$. The  parameter $\sigma$ is randomly generated from an uniform distribution in $[3,10]$. We also replace a certain proportion of observations from each group in the noise variables by uniformly distributed outliers, according to $U[-12,6] \cup U[6,12]$. Note that the observations contaminated in the informative variables differ from those in the noise variables. Furthermore, we always replace (contaminate) the first observations from each group in the informative variables, whereas observations in the noise variables  are randomly selected for the following contamination.

\subsection{Simulation 1: Selection of parameters}

In the first study, we investigate the ability of the modified gap statistic to correctly select the number of clusters $k$ and the sparsity parameter $s$ when applying the introduced algorithm. We consider 100 datasets of 800 dimensions in which the first 50 variables describe the group structure. 
In order to explore the performance of the gap statistic, 3 situations with different numbers of groups are considered, i.e. $g=3, 4, 5$. The sizes of the observations in the groups are randomly selected, ranging from 50 to 150. The contamination strategy corresponds to replacing the first $10\%$ of observations from each group in all informative variables by scattered outliers. In contrast, the uniformly distributed outliers are placed in 75 randomly selected noise variables. 

The proposed method is applied with $k=2,\dots,7$ and various $s$ going from 1.1 up to $\sqrt p$ in steps of 0.5, in order to calculate the gap statistic and to select optimal parameters. 
The results are evaluated  in terms of the estimated number of clusters and the evaluation measures described in Sect. \ref{sec:evalsetup}. It should be noted that CER is calculated with respect to the  group membership before contamination. Since each group is contaminated by scatter outliers, such outliers have the same location as a group and, therefore, they should be assigned to the corresponding group.

\begin{figure}[h]
\begin{center}
    \includegraphics[width=1\textwidth]{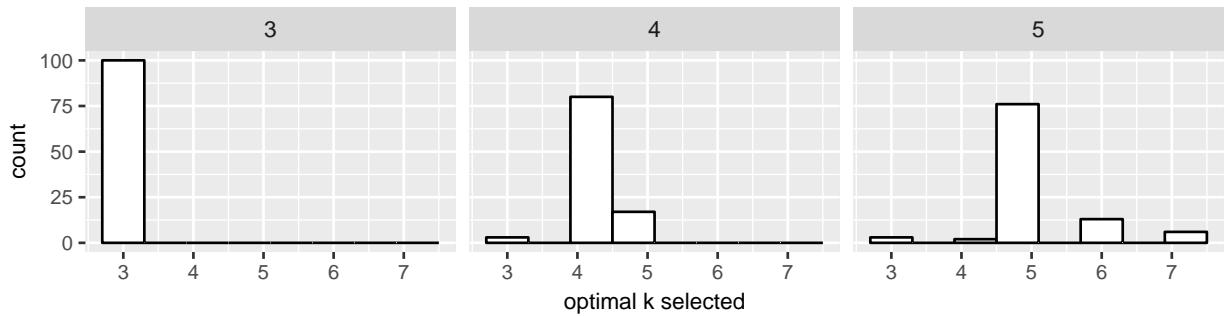}
\caption{Evaluation of the results in terms of the optimal $k$ selected by the gap statistic, for different numbers of groups, i.e. $g=3, 4, 5$. The reported values are based on 100 simulated datasets for each $g$}  \label{Fig7a}
\end{center}
\end{figure}

Figure \ref{Fig7a} summarizes the resulting optimal $k$ selected by the gap statistic as histograms, for the three different numbers of underlying groups ($g=3, 4, 5$). While the gap statistic works perfectly in case of 3 groups, its performance gets slightly worse for a higher number of groups. Nevertheless, the last two histograms clearly indicate that the optimal $k$ is correctly chosen in most cases. 

\begin{figure}[h]
\begin{center}
    \includegraphics[width=1\textwidth]{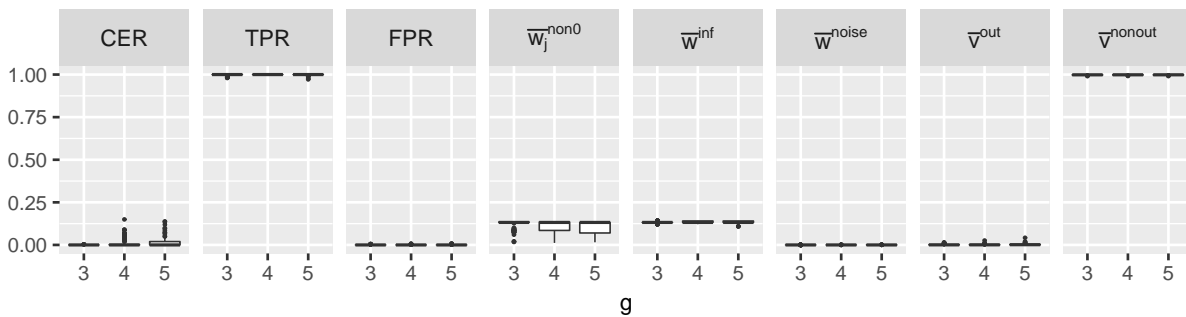}
\caption{Evaluation of the results based on the optimal parameter selection determined by the modified gap statistic, for different numbers of groups, i.e. g=3, 4, 5.
The reported values of the evaluation measures and the selected $k$ represent all 100 simulated datasets} \label{Fig7b}
\end{center}
\end{figure}

Figure \ref{Fig7b} summarizes the results based on the evaluation measures. 
In general, there is no clear dependence between the considered numbers of groups and the resulting values of evaluation measures. Overall, low CER indicate that the proposed procedure can correctly identify the group structure.  In addition, high as well as similar values of  $\bar{w}^{\,inf}$ and $\bar{w}^{\,non0}$ demonstrate the appropriate selection of $s$. Hence, it seems that most of the informative variables can be correctly identified.  The high performance of variable selection is also supported by zero values of  $\bar{w}^{\,noise}$ suggesting that the method is able to discard all noise variables. We also evaluate the method regarding the detection of outliers. We can see that outliers receive on average considerably low weights in contrast to non-outliers; compare $\bar{v}^{out}$ and  $\bar{v}^{nonout}$. Therefore, classifying the observations with $v_i>0.5$ as outliers seems to be a reasonable choice. Indeed, such a cut-off value leads to a great ability to identify outliers indicated by high TPR as well as low FPR. Considering the values of the evaluation measures, we can conclude that the method as well as the parameter selection work efficiently.

\subsection{Simulation 2: Resistance against outliers}

The second simulation study aims at investigating how resistant the proposed method as well as the modified gap statistic are against various proportions of outliers. For this study, we generate 100 datasets that consist of 3 groups of different sizes ranging between 50 and 150 (randomly selected). The data space is defined by 170 informative variables and 830 noise variables, leading to 1000 dimensions in total. Overall, we consider 8 contamination strategies in terms of different percentages of outliers. The datasets in the first strategy are free of outliers. In contrast, the second strategy considers $5\%$ of scatter outliers in all informative variables and no outliers in noise variables. 
The datasets in the remaining strategies are contaminated with $10\%$,  $15\%$, $20\%$, $30\%$, and $40\%$ scatter outliers, respectively, in the informative variables. In addition, the proportion of outliers in the $83$ ($10\%$) noise variables is always kept as $10\%$.

Again, the proposed algorithm is applied with the different numbers of clusters ($k=2, 3, 4, 5, 6$) and various $s$. Subsequently, the gap statistic is employed to estimate the optimal parameter settings. The performance is finally evaluated by the measures described in Sect. \ref{sec:evalsetup} as well as the selected $k$. As in the previous study, we calculate CER by taking the true group membership before contamination into account.

\begin{figure}[h]
\begin{center}
    \includegraphics[width=1\textwidth]{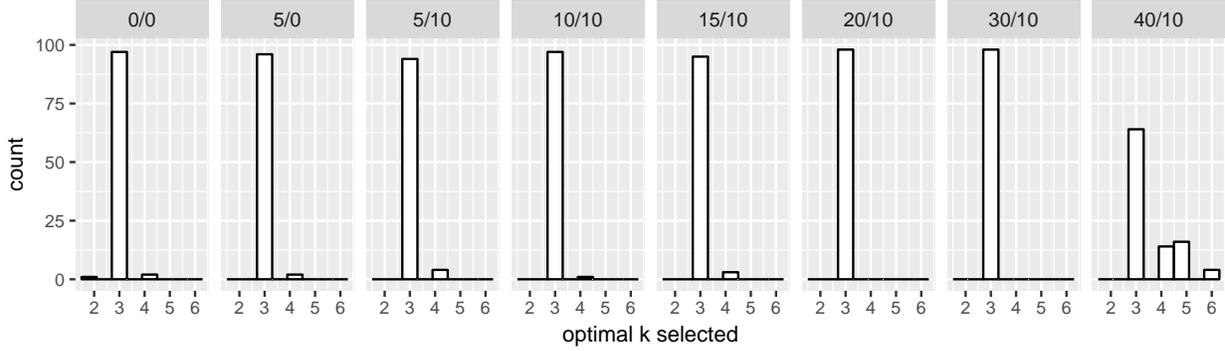}
\caption{Evaluation of the ability to correctly estimate $k$, considering different percentages of outliers in informative and in noise variables (x/x). The reported values of evaluation measures represent all 100 simulated datasets}\label{Fig8a}
\end{center}
\end{figure}

Figure \ref{Fig8a} shows the optimal number of clusters estimated by the proposed gap statistic for each contamination strategy. The histograms clearly indicate that the gap statistic allows to correctly select the number of clusters, i.e. $k=3$, even if data sets contain $40\%$ outliers in total. Although the selection of $k$ seems to be affected by the highest level of contamination corresponding to $50\%$ outliers, such a large proportion of outliers is however very extreme and unrealistic in practice.

\begin{figure}[h]
\begin{center}
    \includegraphics[width=1\textwidth]{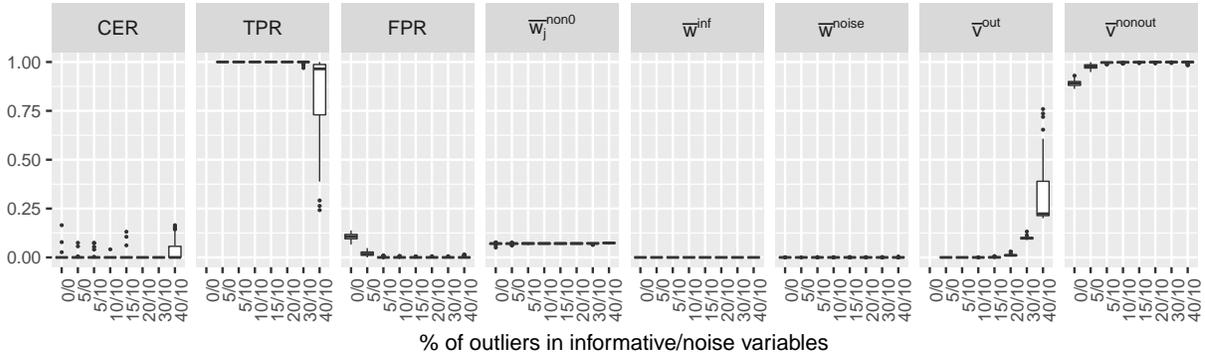}
\caption{Evaluation of the results considering different percentages of outliers in informative and in noise variables (x/x). The reported values of evaluation measures represent all 100 simulated datasets} \label{Fig8b}
\end{center}
\end{figure}

Figure \ref{Fig8b} summarizes the performance in terms of evaluation measures and demonstrates a great ability to discover the group structure independently of the number of outliers, reflected by low CER. The low CER can also be observed in case of the highest contamination. This might indicate that even if the gap statistic estimates a higher number of clusters than the actual number of groups (see Figure \ref{Fig8a}), the detected clusters seem to be to some extent still homogeneous. The great performance of the gap statistic is additionally supported by similar values of $\bar{w}^{non0}$ and $\bar{w}^{inf}$, indicating highly efficient variable selection. Furthermore, zero values of $\bar{w}^{noise}$  imply that most noise variables are discarded for data clustering. Therefore, we can assume that the sparsity parameter $s$ is appropriately estimated.
Regarding the detection of outliers, the method can identify most outliers indicated by TPR around 1. 
However, TPR is slightly below 1 for the extremely contaminated data sets (i.e. 40/10). Such low TPR can be a consequence of high observation weights for outliers, reflected by higher $\bar{v}^{out}$. This might indicate that the weights of some outliers are similar to the weights of non-outliers, or the cut-off value $0.5$ needs to be increased in order to achieve perfect outlier detection for a large contamination level. Although the method misclassifies around 10\% of normal observations in case of no contamination (0/0), it is able to correctly classify almost all non-outliers in contaminated datasets indicated  by zero FPR. Based on the overall evaluation, the method demonstrates a great ability to identify a complex data structure in contaminated datasets.

\subsection{Simulation 3: Comparison }

In the last study, we compared the proposed weighted robust and sparse $k$-means (WRSK) algorithm with other $k$-means-based approaches, such as $k$-means (KC), trimmed $k$-means (TKC), sparse $k$-means (SKC) and its trimmed version (RSKC) on 30 simulated datasets. Each dataset is represented by 4 groups of various sizes ranging between $15$ and $150$. The generated observations are described by 4000 variables. Since the additional goal is to investigate the influence of different proportions of informative variables, three settings are considered, such as a percentage of 1\%, 2\%, and 5\% of informative variables. Moreover, $20\%$ of the observations are replaced by uniformly distributed outliers in the first $20\%$ of the informative variables, and $10\%$ of other observations are contaminated in $20\%$ of randomly selected noise variables.

When applying the methods on the generated datasets, we assume prior knowledge of the number of clusters and optimize $s$ in case of sparse-based algorithms. The trimming level for both TKC and RSKC corresponds to the total percentage of outliers, i.e. $\alpha=0.30$. We evaluate the clustering solution by CER, and if appropriate, the performance regarding the outlier detection by TPR and FPR. Note that CER is again calculated with respect to the true group memberships before contamination. Since the outliers are placed only in the subset of informative variables, there is still some information about the group separation in non-contaminated variables. 

\begin{figure}[h]
\begin{center}
    \includegraphics[width=1\textwidth]{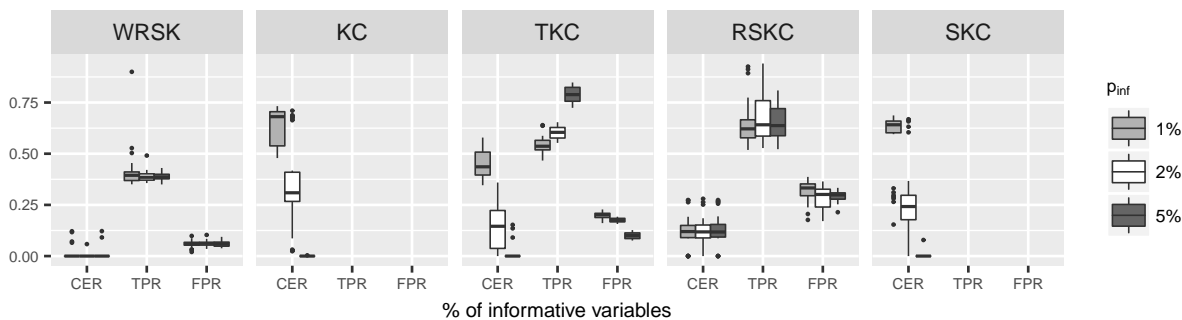}
\caption{Evaluation of various $k$-means based clustering methods, considering various proportions of informative variables ($p_{inf}$). The reported values of the evaluation measures correspond to the 30 simulated datasets} \label{Fig9}
\end{center}
\end{figure}

Figure \ref{Fig9} summarizes the result based on 30 simulations. In general, in comparison to the remaining $k$-means-based methods, both the proposed method and RSKC seem to be resistent against the different percentages of informative variables. The clustering performance of KC, TKC, and SKC increases with an increasing proportion of informative variables, indicated by decreasing CER. In addition, CER shows that the proposed method outperforms the remaining methods in terms of identifying the underlying group structure reflected by the lowest CER for most simulated datasets. Although lower TPR demonstrate that our WRSK is not capable of identifying all outliers in comparison to the trimmed-based methods, the proposed method misclassifies fewer non-outliers indicated by the lowest FPR. Considering the performance, it seems that our method is able to sufficiently identify the group structure even if a large amount of noise variables is present in a data set.

\section{Analyzing the group structure of glass vessels} \label{sec:application}

\begin{figure}[h]
\begin{center}
    \includegraphics[width=0.5\textwidth]{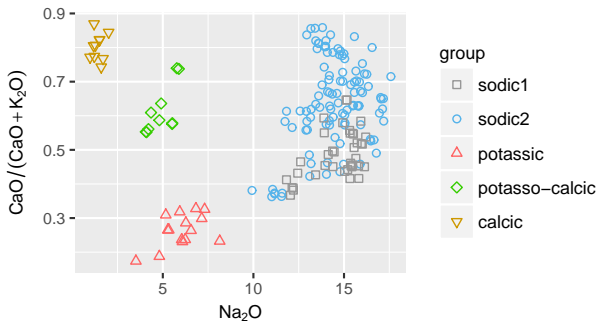}
\caption{Group membership of analyzed glass vessels, based on element concentrations \citep{lemberge2000quantitative}  }\label{fig:membership}
\end{center}
\end{figure}

The proposed algorithm is particularly useful in the situation where a large number of variables is present as in the case of archaeological glass vessels from the $16^{th}$  and $17^{th}$ centuries, which were excavated in Antwerp being one of the most important historical centers of both glass manufacturing and trade.  In 1997, chemical analysis was conducted in order to get better insight into the glass collection, including also the possible origin of the various glass samples. For this reason, the glass vessels were analyzed by an electron-probe X-ray micro-analysis (EXPMA) to measure spectra at different energy levels \citep{Janssens1998}. Consequently, traditional calibration methods were applied on spectra to extract major chemical elements resulting in the separation of four glass vessels groups, i.e. sodic, potasso-calcic, calcic, and potassic.  The connection between element concentrations of glass vessels and their origin was discussed by \citet{Janssens1998}. \citet{lemberge2000quantitative} used their findings on an extended dataset consisting of 180 glass samples described by 1920 variables (different energy levels) in order to predict the same concentrations of the major elements as \citet{Janssens1998}, using partial least squares.   In this paper we employ the extended dataset consisting of 4 groups as well, as shown in Fig. \ref{fig:membership}. The plot additionally shows that the largest group (sodic) is split into two subgroups that are not clearly separated in the two-dimensional space of chemical concentrations. The two subgroups are caused by the installation of  different detector efficiencies in the EXPMA. Detecting the subgroup of glass samples analyzed after the installation has been investigated e.g. by \citet{serneels2005partial} and \citet{Filzmoser2008}.

Our focus is to detect an entire group structure, i.e. 5 groups, which might be hidden in the high-dimensional data space. Note that there is no pre-knowledge about the informative variables, neither of outliers in each group. In addition, the group membership based on the chemical concentrations does not necessarily have to reflect the group structure based on the origin of the glass samples. However, there exist some assumptions about the connection between the chemical elements - the glass manufacturing process - and the origin \citep{Janssens1998}.  Therefore, we evaluate the performance in terms of CER with respect to the group membership shown in Fig. \ref{fig:membership}, and the cluster membership obtained by k-means-based algorithms with $k=5$. Although it is not sure whether or not the dataset contains outliers, we set the trimming level to 0.10 for the trimming-based methods as suggested by \citet{Kondo2016}. The optimal sparsity parameter for RSKC and WRSKC is selected from 1.5 to $\sqrt{p}$ in steps of 0.1 based on the gap statistic described in Sect.~\ref{sec:parameters}. The evaluation of the resulting clustering solution is presented in Table \ref{tab:cer}, which clearly shows that WRSK outperforms the remaining methods indicated by the lowest CER. Incorporating the trimming concept or sparsity seem to improve the performance of $k$-means (KC) as demonstrated by slightly larger CER for TKC or SKC. RSKC shows the worst performance. The reason might be that either important variables have been excluded, or that wrong observations have been trimmed, or a combination of both.

\begin{table}
\begin{center}
\caption{Evaluation of the clustering performance of $k$-means-based clustering methods}
\label{tab:cer}       
\begin{tabular}{llllll}
\hline
method &  WRSK & KC & TKC & RSKC & SKC\\
\hline
CER & $0.039$ & $ 0.183$  & $0.166$   & $0.191$   & $0.167$\\
\hline
\end{tabular}
\end{center}
\end{table}

\begin{figure}
\begin{center}
    \includegraphics[width=1\textwidth]{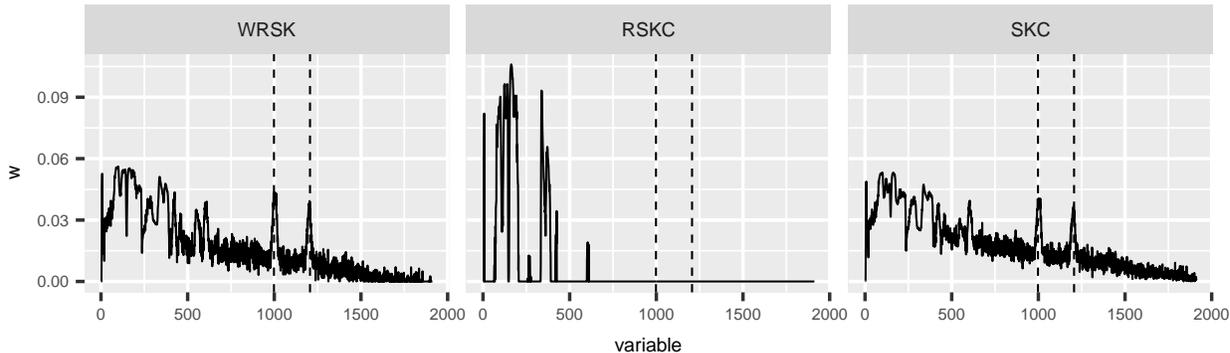}
\caption{The final variable weights obtained by sparse $k$-means-based clustering methods}\label{fig:glassweights}
\end{center}
\end{figure}

We also examine the final variable weights obtained by the sparse k-means-based algorithms. Figure \ref{fig:glassweights} shows the final weights for each sparse method. The resulting values of the weights demonstrate that SKC completely fails in terms of achieving sparsity in the variable weight vector, as $w_j>0$ for almost all variables. Nevertheless, there are several variables that receive a higher weight than in case of RSKC; see two peeks highlighted by dashed lines. This may indicate that there could be useful information about the group separation in the last energy levels of the measured spectra. A very similar conclusion can be made for the weights obtained by the proposed WRSK. In addition, WRSK results in a slightly sparse variable weight vector and at the same time can appropriately identify 5 groups as indicated by the lowest CER.

\begin{figure}
\begin{center}
    \includegraphics[width=1\textwidth]{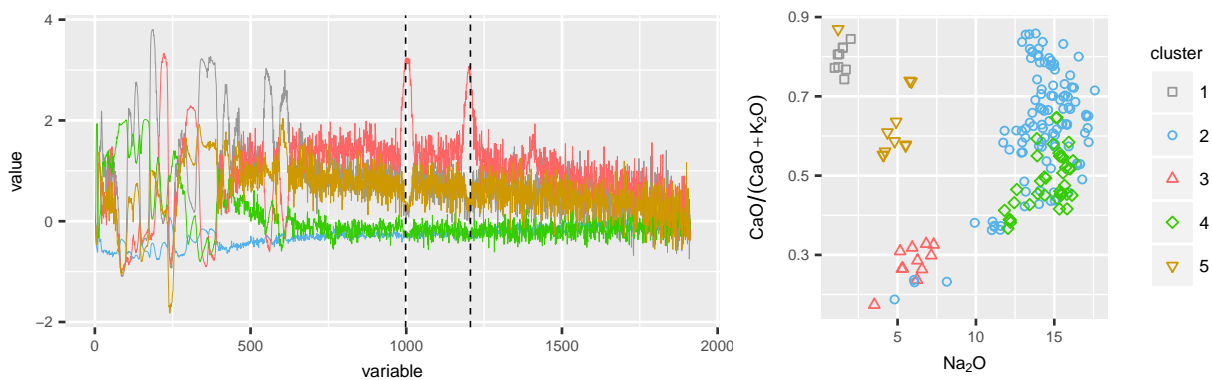}
\caption{Cluster centers calculated for each variable with respect to the final cluster membership obtained by WRSK (left) and the corresponding cluster membership (right)}\label{fig:ourclusters}
\end{center}
\end{figure}

In order to investigate the final variable weights obtained by the proposed method in more detail, we examine how the centers of the detected clusters are distinguishable at each energy level of the spectra, i.e. for each variable. For this reason, we calculate the cluster centers  as a weighted mean of the observations in each variable with the corresponding observation weights and the identified cluster membership.  The resulting centers are displayed as spectra in Fig \ref{fig:ourclusters} (left) and are distinguished by different colors based on the final cluster membership visualized in Fig \ref{fig:ourclusters} (right).  Figure \ref{fig:ourclusters} (left) particularly indicates that the centers appear to be well separated already at the low energy levels, i.e. in the first part of the variable vector.  Furthermore, the center of cluster 3 appears to be well separated from other centers in the higher energy levels, highlighted by two dashed lines. 
In fact, the proposed WRSK is capable of identifying this informative part of the spectra; see Fig.~\ref{fig:glassweights}. Although the proposed method does not lead to high sparsity in the variable weight vector, the final cluster membership visualized in Fig.~\ref{fig:ourclusters} (right) indicates a great ability of WRSK to correctly identify informative variables since all 5 glass vessels groups are well recovered with only 5 misclassified observations. Whereas the misclassified calcic glass sample has an observation weight equal to 1, the remaining four misclassified potassic glass samples obtain weights considerably smaller than 1, i.e. $0.06, 0.00, 0.60, 0.76$. This might indicate that although these observations are originally from the potassic group, their chemical structure seems to be different from the remaining observations of that group.

\section{Conclusion} \label{sec:discussion}

We propose a $k$-means-based clustering procedure that endeavors to simultaneously detect groups, outliers, and informative variables in high-dimensional data. The motivation behind our method is to improve the performance of the popular $k$-means method for real-world data that possibly contain both outliers and noise variables.  \citet{Kondo2016} have addressed both issues in the robust (trimmed) and sparse $k$-means procedure, but our method goes even further. Firstly, our method aims to identify
clusters, outliers, and noise variables at the same time. Secondly, the proposed procedure is designed in such a way that the required parameters are automatically estimated and, therefore, no pre-knowledge about the data is required.
By incorporating the weighting function in $k$-means, each observation automatically receives a weight reflecting the degree of outlyingness based on which the outliers are identified. In order to correctly detect the informative variables, we employ a sparsity concept adjusted by observation weights. The proposed modified gap statistic is employed to optimize both the sparsity parameter and the number of clusters. 

The introduced method together with the modified gap statistic has thoroughly been tested on a variety of simulated data sets as well as on a  high-dimensional real data set. The conducted experiments indicated a great ability of the proposed procedure to discover the group structure. The presented clustering algorithm as well as the data generating processes are implemented in the R package \texttt{wrsk}, freely available at \texttt{https://github.com/brodsa/wrsk}. 

Future research includes extending the analysis of a data structure to identify the variables which are responsible for outliers.  Such an idea is closely related to cell-wise outlier detection by \citet{Rousseeuw2016}  for the situation of a
single group data structure. A similar concept was introduced by \citet{farcomeni2014snipping} in the context of clustering. The aim was to demonstrate that cell-wise contamination does not affect the introduced approach. However, the method has been tested in terms of clustering only, and no investigation has been conducted with respect to cell-wise outlier detection. Considering that outliers are commonly highly interesting observations due to their typically different content, it is even more important to find out which variables are behind this unusual behavior.

\acknowledgments 
This work has been partly funded by the Vienna Science and Technology Fund (WWTF) through project ICT12-010 and by the K-project DEXHELPP through COMET - Competence Centers for Excellent Technologies, supported by BMVIT, BMWFW and the province Vienna. The COMET program is administrated by FFG.

%
%

\end{document}